\documentclass[sigconf]{acmart}
\usepackage{textcomp}
\usepackage{xcolor}
\usepackage{soul}
\usepackage{braket}
\usepackage{algorithmic}
\usepackage[ruled,vlined, linesnumbered]{algorithm2e}
\usepackage{graphicx}
\AtBeginDocument{%
  \providecommand\BibTeX{{%
    \normalfont B\kern-0.5em{\scshape i\kern-0.25em b}\kern-0.8em\TeX}}}

\def\BibTeX{{\rm B\kern-.05em{\sc i\kern-.025em b}\kern-.08em
    T\kern-.1667em\lower.7ex\hbox{E}\kern-.125emX}}
\begin{document}
\fancyhead{}
\title{Robust and Secure Hybrid Quantum-Classical Computation on Untrusted Cloud-Based Quantum Hardware}


\author{Suryansh Upadhyay}
\affiliation{%
  \institution{The Pennsylvania State University}
  \streetaddress{University Park}
  \city{PA}
  \country{USA}}
\email{sju5079@psu.edu}

\author{Swaroop Ghosh}
\affiliation{%
  \institution{The Pennsylvania State University}
  \streetaddress{University Park}
  \city{PA}
  \country{USA}}
\email{szg212@psu.edu}


\begin{abstract}
Quantum computers are currently accessible through a cloud-based platform that allows users to run their programs on a suite of quantum hardware. As the quantum computing ecosystem grows in popularity and utility, it is reasonable to expect more companies, including untrustworthy/untrustworthy/unreliable vendors, to begin offering quantum computers as hardware-as-a-service at various price/performance points. Since computing time on quantum hardware is expensive and the access queue may be long, users will be enticed to use less expensive but less reliable/trustworthy hardware. Less-trusted vendors may tamper with the results and/or parameters of quantum circuits, providing the user with a sub-optimal solution or incurring a cost of higher iterations. In this paper, we model and simulate adversarial tampering of input parameters and measurement outcomes on an exemplary hybrid quantum classical algorithm namely, Quantum Approximate Optimization Algorithm (QAOA). We observe a maximum performance degradation of $\approx 40\%$. To achieve comparable performance with minimal parameter tampering, the user incurs a minimum cost of 20X higher iteration. We propose distributing the computation (iterations) equally among the various hardware options to ensure trustworthy computing for a mix of trusted and untrusted hardware. In the chosen performance metrics, we observe a maximum improvement of $\approx$30\%. In addition, we propose re-initialization of the parameters after a few initial iterations to fully recover the original program performance and an intelligent run adaptive split heuristic, which allows users to identify tampered/untrustworthy hardware at runtime and allocate more iterations to the reliable hardware, resulting in a maximum improvement of $\approx$$45\%$.

\end{abstract}





\maketitle

\section{Introduction}

 Quantum computing (QC) can solve many combinatorial problems exponentially faster than classical counterparts by leveraging superposition and entanglement properties. Example include machine learning \cite{b1}, security \cite{b2}, drug discovery \cite{b3}, computational quantum chemistry \cite{b4} and optimization \cite{b5}. Since noisy computers are less powerful and qubit limited, various hybrid algorithms are being pursued, such as, the Quantum Approximate Optimization Algorithm (QAOA) and Variational Quantum Eigensolver (VQE), in which a classical computer iteratively drives the parameters of a quantum circuit. The purpose of the classical computer is to tune the parameters that will guide the quantum program to the best solution for a given problem. On high-quality hardware with stable qubits, the algorithm is likely to converge to the optimal solution faster, i.e., with fewer iterations. Currently, users can create circuits for specific hardware and upload them to the cloud, where they are queued. The results of the experiment are returned to the user once the experiment is completed. However, with the current setup, the user has little/no visibility into the hardware assigned to the program. The hardware vendors of quantum computers provide a compiler for their hardware, such as IBM's Qiskit compiler \cite{b6}, Rigetti's QuilC compiler \cite{b7}, and so on. Furthermore, some third-party software tools, such as, compilers like Orquestra \cite{b8} and tKet \cite{b9}, are appearing that support hardware from multiple vendors. As the quantum computing ecosystem evolves, more third-party service providers will emerge offering potentially higher performance. This will entice users to utilize these services. 
 
\begin{figure}
    \centering
    \includegraphics[width= 3.4in]{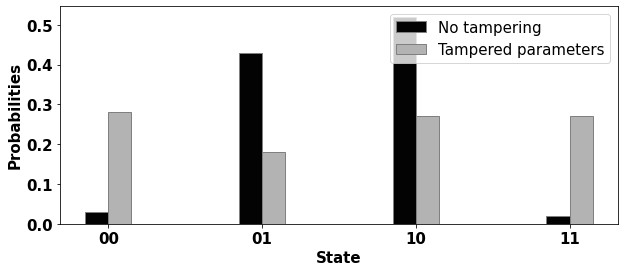}
    \caption{Sample 2-node maxcut using QAOA (correct output = `10' or `01'), simulated on the fake back-end (Fake$_-$montreal) for 20 iterations. States `10' or `01' are no longer the dominant output with adversarial tampering.}
    \label{1}
\vspace{-5mm}
\end{figure}

\textbf{Proposed attack model:} In this paper, we discuss a security risk associated with the use of third-party service providers and/or any untrusted vendor for hybrid quantum classical workloads (QAOA). In the proposed attack model, less-trusted quantum service providers can pose as trustworthy hardware providers and tamper with the results and/or input training parameters, resulting in users receiving a sub-optimal solution or incurring a cost of higher iterations. To show the extent of damage done by tampering, we run a simple 2-node maxcut using QAOA on tampered (parameter tampering with x=25\%, where x is the tampering error in parameters) and non-tampered hardware and compare the probability distributions of basis states for both cases (Fig. \ref{1}). 
The correct output is `10' or `01'. With adversarial tampering, states `10' or `01' are no longer the dominant output for the same number of iterations. In practical scenario, the user must rely on the sub-optimal output of the tampered quantum computer as the correct output of the optimization problem is unknown. 

\begin{figure}
    \centering
    \includegraphics[width= 3.4in]{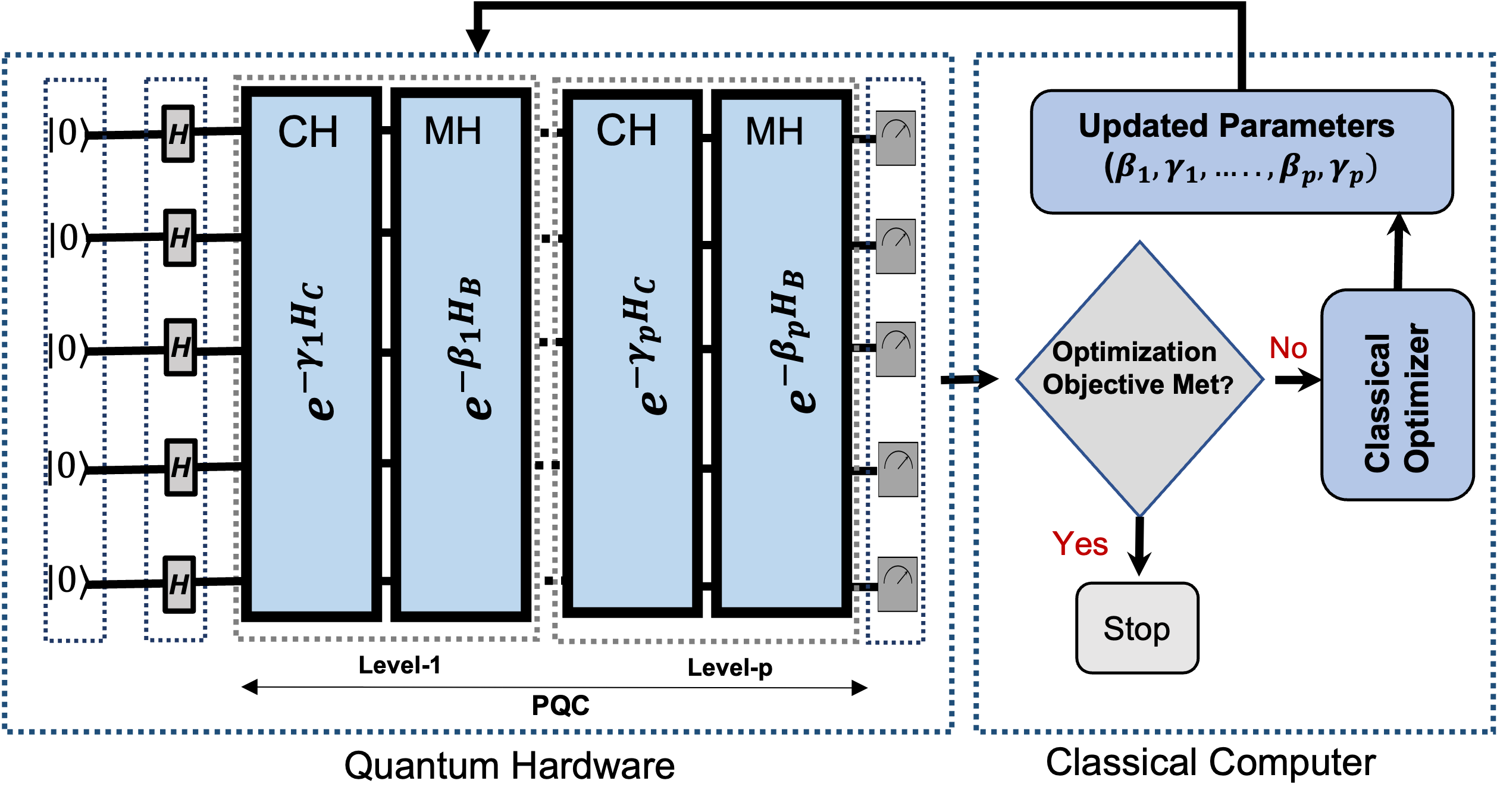}
    \caption{Schematic of a p-level quantum-classical hybrid algorithm QAOA. A quantum circuit takes input qubit states and alternately applies Cost Hamiltonian (CH) and Mixing Hamiltonian (MH) `p' times and the final state is measured to obtain expectation value with respect to the objective function. A classical optimizer finds the best parameters$(\gamma, \beta)$ that maximizes the cost.}
    \label{2}
\vspace{-5mm}
\end{figure}

\textbf{Proposed solution:} To mitigate the adversarial tampering we propose, (a) splitting the number of iterations on multiple available hardware. The idea is to distribute the computation among the various hardware (a mix of trusted and untrusted ones or mixture of untrusted hardware from multiple vendors) available. The results from individual hardware and iterations are stitched or combined to obtain the probability distribution of the solution space, (b) an intelligent run-adaptive iteration distribution to identify untrusted hardware and bias the number of iterations to favor trusted/reliable hardware to maximize the overall computation quality. The motivation is that the users may end up using trusted and untrusted hardware equally which may not be optimal in terms of performance, and (c) re-initialization of the parameters after a few initial iterations to partially or completely recover the original program performance against an attack in which the adversary introduces significant error in the parameters during the first few iterations. 

\textbf{Contributions:} To the best of our knowledge, this is the first effort on considering adversarial tampering on hybrid quantum-classical algorithms. In particular, we (a) propose a model for result tampering and parameter tampering for untrustworthy third-party hardware vendors, (b) demonstrate the effectiveness of our proposed attack model using  hybrid quantum-classical algorithm QAOA on a fake back-end, (c) propose equally distributing computation among the various available hardware, intelligent run-adaptive iteration distribution, and re-initialization of the parameters after a few initial iterations for resilience, (d) assess and validate our proposed defense.

\textbf{Related work:} Several recent works on the security of quantum computing \cite{b11, b12, b13, b14} exist in literature. The authors of \cite{b13} consider an attack model where a rogue element in the quantum cloud reports incorrect device calibration data, causing a user to run his/her program on an inferior set of qubits. The authors propose that test points be added to the circuit to detect any dynamic malicious changes to the calibration data. The objective of our attack model is to tamper with the result/parameters so that incorrect or sub-optimal outcome is reported to the user. In \cite{b11}, Ensemble of Diverse Mappings (EDM) is proposed to tolerate correlated errors. However, this is yet another case of mapping agnostic optimization. Using dummy gates, a closely related work \cite{b12} conceals the functionality of a quantum circuit from untrusted compilers. Another work \cite{b14} proposes split compilation to address potential security issues caused by untrusted third-party compilers. The proposed attack model is very different than the untrusted compiler where the objective is to steal the intellectual property embedded within the quantum circuit.

The remainder of the paper is structured as follows: Section II provides background on various concepts in quantum computing relevant to this work. The proposed attack model is described in Section III. Section IV presents the simulation results and analysis. Section V proposes and evaluates the proposed defense using simulations and experiments. Section VI discusses about other possible attack scenarios and assumptions. Section VII concludes the paper.
\vspace{-1mm}
\section{Background}

\subsection{Qubits and Quantum Gates}

Qubits are the building blocks of a quantum computer that store data as various internal states (i.e., $\ket{0}$ and $\ket{1}$).A qubit state is represented as  $\varphi$ = a $\ket{0}$ + b $\ket{1}$ where a and b are complex probability amplitudes of states $\ket{0}$ and $\ket{1}$ respectively.The gate operations change the amplitudes of the qubits to produce the desired output. A quantum program executes a series of gate operations (using laser pulses in ion trap qubits and RF pulses in superconducting qubits) on a group of correctly initialized qubits.  
\vspace{-1mm}
\subsection{Quantum Error}

Quantum gates are realized with pulses that can be erroneous. Quantum gates are also prone to error due to noise and decoherence. The deeper quantum circuit needs more time to execute and gets affected by decoherence which is usually characterized by the relaxation time (T1) and the dephasing time (T2). The buildup of gate error is also accelerated by more gates in the circuit. Because of measurement circuitry imperfections, reading out a qubit containing a 1 may result in a 0 and vice versa. Unlike gate-error and decoherence, which depend on the number of gates in the circuit, readout error is gate count agnostic. It solely depends on the state being read.
\vspace{-2mm}
\subsection{Cloud-based Quantum Backends}

Some of the leading companies currently offering users access to quantum hardware via the cloud include IBM, Google, Microsoft, Qutech, QC Ware, and AWS Braket (both superconducting and Trapped Ion qubits). Users can create and modify quantum applications with the help of toolkits known as quantum software development kits. Although using a quantum computer is expensive, cloud-based access encourages users to look into less expensive quantum hardware instead. In the future, less reputable third parties might sell quantum computers (that may be located offshore). This situation is very similar to untrusted compilers that may produce very optimal circuit but may steal sensitive intellectual property \cite{b12}. 
\vspace{-2mm}

\begin{figure*}
    \centering
    \includegraphics[width= 7in]{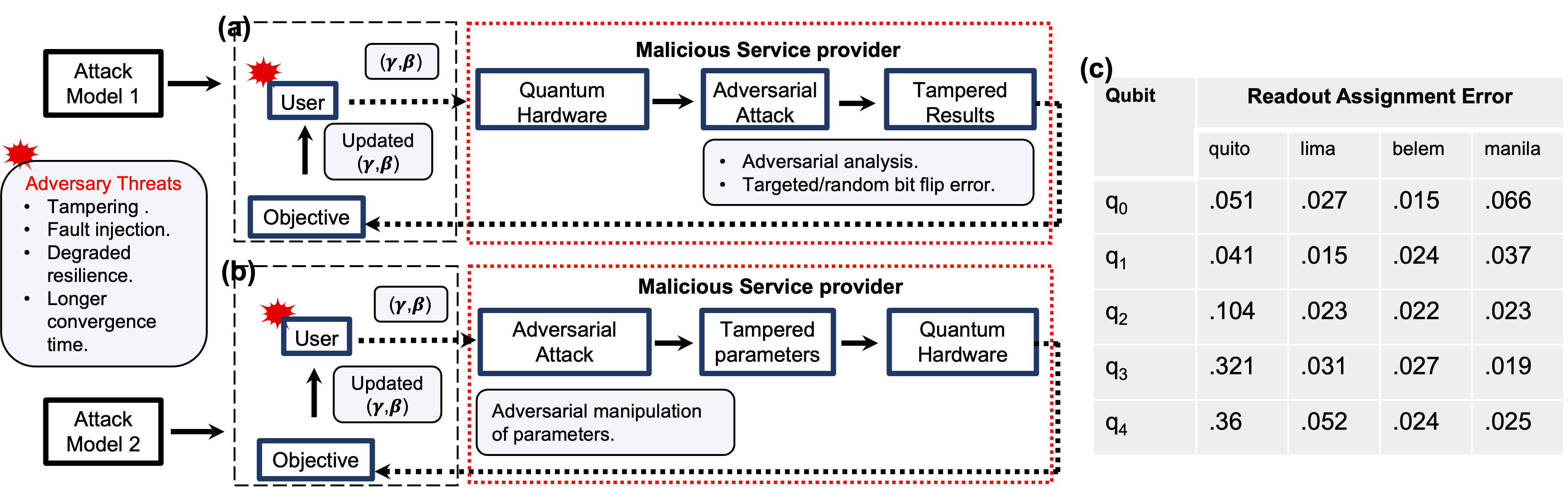}
    \caption{Proposed attack models. a) Attack model1 considers an adversary who tampers with the outcome after an iterative program has been run on the hardware. b) In Attack model 2, the adversary tampers with the parameters before running the program on a quantum computer. c) Readout assignment errors for various IBM quantum hardware.}
    \label{3}
\vspace{-2mm}
\end{figure*}

\subsection{Parameterized quantum circuit(PQC)}

A collection of parameterized and parametrized single qubit gates make up PQC. The parameters are iteratively optimized by a classical optimizer to produce the desired input-output relationship. A quantum processor creates a quantum state using a PQC. The output distribution obtained by repeatedly measuring the quantum state is then fed to a classical optimizer. The classical computer creates a fresh set of optimized PQC parameters (using well-known optimizers such as, gradient descent) based on the output distribution, which is then fed back to the quantum computer. Up until the optimization goal is achieved, the entire process is kept running in a closed loop. The number of iterations to converge to a solution is an important aspect of the PQC. One can randomly initialize the parameters or employ intelligent approaches to initialize the parameters to accelerate convergence. Longer convergence time can be expensive since user has to wait longer and pay more due to higher usage of the quantum resources. In recent years, quantum routines that are inherently resilient to errors have been developed using PQC.  
\vspace{-1mm}
\subsection{Quantum Approximate Optimization Algorithm(QAOA)}

QAOA is a hybrid quantum-classical variational algorithm designed to tackle combinatorial optimization problems. A p-level variational circuit with 2p variational parameters creates the quantum state in QAOA. Even at the smallest circuit depth (p = 1), QAOA delivers non-trivial verifiable performance guarantees, and the performance is anticipated to get better as the p-value increases \cite{b5}. Fig. \ref{2} shows an overview of QAOA to solve a combinatorial problem.  Recent developments in finding effective parameters for QAOA have been developed are reported in \cite{b5,b15,b16}. In QAOA, a qubit is used to represent each of the binary variables in the target C(z). In each of the p levels of the QAOA circuit, the classical objective function C(z) is transformed into a quantum problem Hamiltonian (Fig.\ref{2}). With optimal values of the control parameters, the output of the QAOA instance is sampled many times and the classical cost function is evaluated with each of these samples. The sample measurement that gives the highest cost is taken as the solution \cite{b17}. In a quantum classical optimization procedure,the expectation value of $H_C$ is determined in the variational quantum state $ E_p(\gamma,\beta)= \varphi_p(\gamma, \beta)|H_C|\varphi_p(\gamma, \beta)$. A classical optimizer iteratively updates these variables $(\gamma, \beta)$ so as to maximize $E_p(\gamma, \beta)$. A figure of merit (FOM) for benchmarking the performance of QAOA is the approximation ratio (AR) and is given as \cite{b18}
\begin{equation}\label{eq:AR}
 AR = E_p(\gamma, \beta)/Cmax
\end{equation}
where $Cmax = MaxSat(C(z))$.

\section{Proposed attack model}

\subsection{Basic Idea}

We consider that the quantum hardware available via cloud service may tamper with the parameters of the input PQC and/or the computation outcome. The objective is to manipulate the results that could have financial and/or socio-political implications. This is feasible since in the future, (a) an untrusted third party may offer access to reliable and trusted quantum computers, such as those from IBM, but may tamper with the circuit/computation results, (b) an untrusted vendor may offer access to untrusted quantum hardware via cloud at a lower price and/or faster access (without a wait queue), incentivizing users to avail their services. Less-trusted quantum service providers can masquerade as trustworthy hardware providers and inject targeted/random tampering, resulting in a suboptimal solution being returned to users. For both scenarios, the user will be forced to trust the less-than-ideal output from the quantum computer since the correct solution to the optimization problem is unknown. The proposed attack models are depicted in Fig. \ref{3}. We propose two distinct attack strategies. Model-1 in Fig.\ref{3}a) involves the adversary tampering with the results while measuring, whereas Model-2 in Fig.\ref{3}b) involves the adversary tampering with the parameters prior to running the program on the quantum hardware.

\begin{figure}
    \centering
    \includegraphics[width= 3.4in]{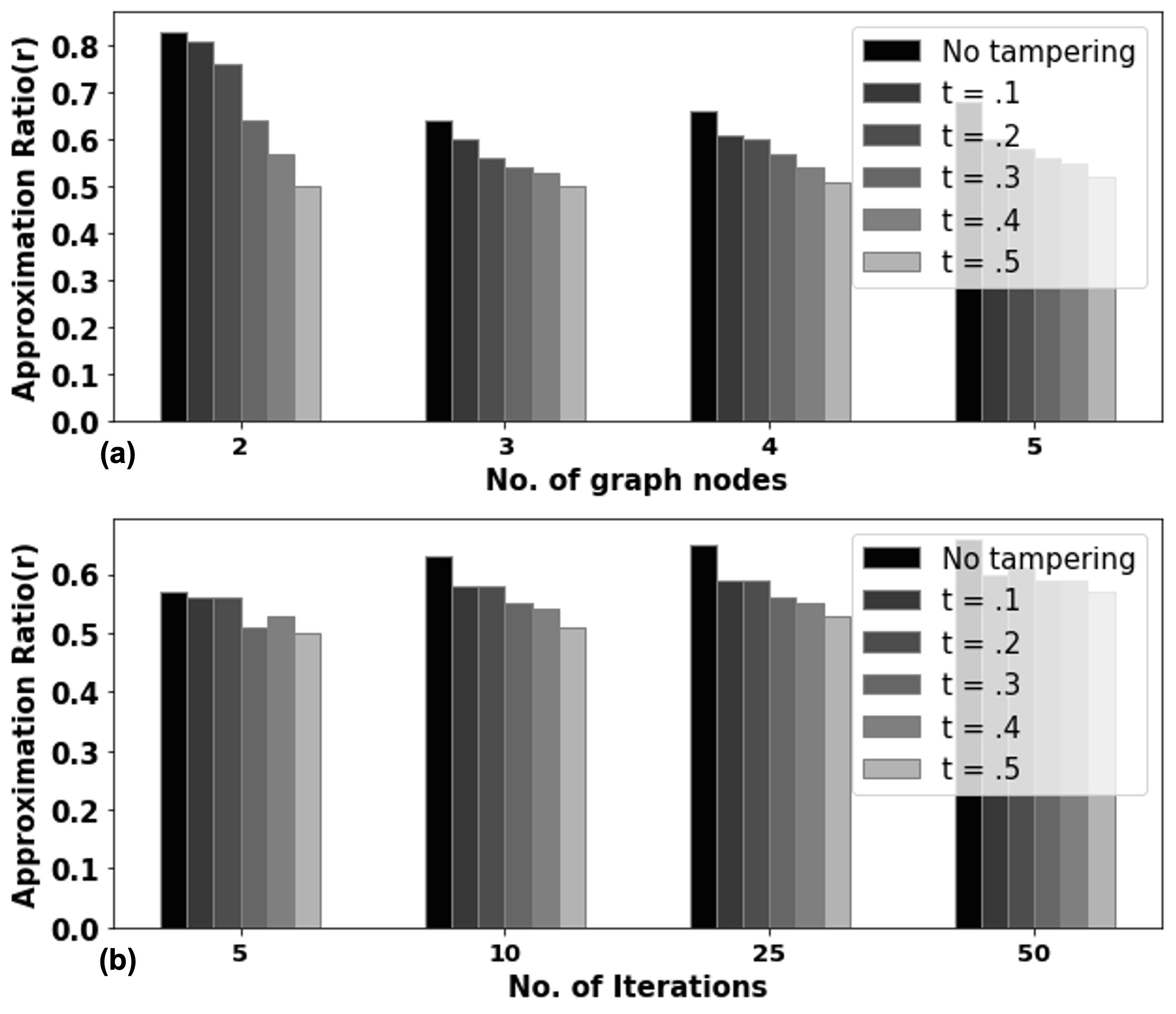}
    \caption{a) Approximation ratio (r) variation for different graph sizes when run on tampered fake$_-$montreal hardware with results for 50 iterations, b) performance comparison of QAOA for four-node graph with varying iterations and tampering on fake$_-$montreal backend.}
    \label{4}
\vspace{-4mm}
\end{figure}

\subsection{Adversary Capabilities}

For the sake of simplicity, we assume that the adversary, (a) has access to either measured results or trainable parameters of the program run by the user. This is likely if the quantum computing cloud provider is rogue, (b) does not manipulate the structure of the quantum circuit. This is possible since tampering the quantum circuit may drastically alter the mapping and other constraints and the computation outcome can be fully corrupted which can be suspected, (c) has the computational resources to analyze the program results to determine which qubit lines or parameters to tamper with, (d) can tamper the parameters (all or few) and/or outputs (all or few) for all iterations or a fraction of iterations.

\subsection{Adversarial Tampering Model}

In the proposed attack model (Fig. \ref{3}), the adversary takes the form of a less reliable/untrusted quantum service provider posing as a reliable/trusted hardware provider. The adversary then tampers with the solution, either by modifying it before reporting it to the user or by modifying the user input parameters before running the iterative circuit on the quantum hardware. It should be noted that the adversary has access to computation results (e.g., qubit basis state probabilities or trainable parameters) before sending them to the user. Assuming that the quantum circuit is correct and optimal, the adversary's solution will be optimal, which can then be tampered with. Various rogue providers may adopt their favorite method for tampering the results. Some examples are as follows.

\subsubsection{Tampering of results}

While measuring the qubit lines, the adversary can introduce random bit flip error (Fig. \ref{3} attack model 1) which we quantify in the paper as the \emph{tampering coefficient}. The adversary can introduce tampering in the form of qubit measurement error, either on all of the qubit lines or on a subset of the qubit lines randomly or the attack can be more strategic in nature focusing on specific qubit lines to introduce measurement errors. For the sake of simplicity, we consider the case where the adversary introduces a bit-flip error in only one of the qubit lines at random. When this extra bit-flip measurement error is added to the readout error, the readout error does not change significantly. For example : 

Example 1 : We consider the real hardware measurement errors, quoted as Readout assignment error (RAE), for the IBM's 5 qubit devices (Fig. \ref{3}(c)). Assuming tampering and RAE to be independent and uncorrelated sources of error, we can get the total error as:

\begin{equation}\label{Equation 2}
    \Delta RAE_{net}  = \sqrt{(\Delta RAE_{qi})^2 + (\Delta Tampering)^2 }
\end{equation}
where $(\Delta RAE_{qi}) = RAE$ value for $i^{th}$ qubit line and $\Delta Tampering$ is defined as :
\begin{equation}\label{eq:eq 3}
    \Delta Tampering  = t/n 
\end{equation}
where $t=$ tampering coefficient,
$n=$ (total qubit lines$ - $tampered qubit lines + 1).

Assuming that the adversary uses ($t =0.1$ or $t =0.5$) , for ibmq$_-$lima $q_1$ RAE, we can calculate the final measurement error (which is 0.028 and 0.09, respectively) for that qubit line using \emph{equation} 2 for targeted tampering. These final error values are comparable to the values quoted for various devices Fig. \ref{3}c and qubit lines. For example the new RAE ($t =0.1$) for ibmq$_-$lima $q_1$ is comparable to RAE's of $q_0$, $q_2$ and even less than $q_4$.  When the RAE value for $t =0.5$ is compared with the tamper-free RAE values of other hardware such as ibmq$_-$quito $q_3$, it is still found to be less.

\begin{figure}
    \centering
    \includegraphics[width= 3.4in]{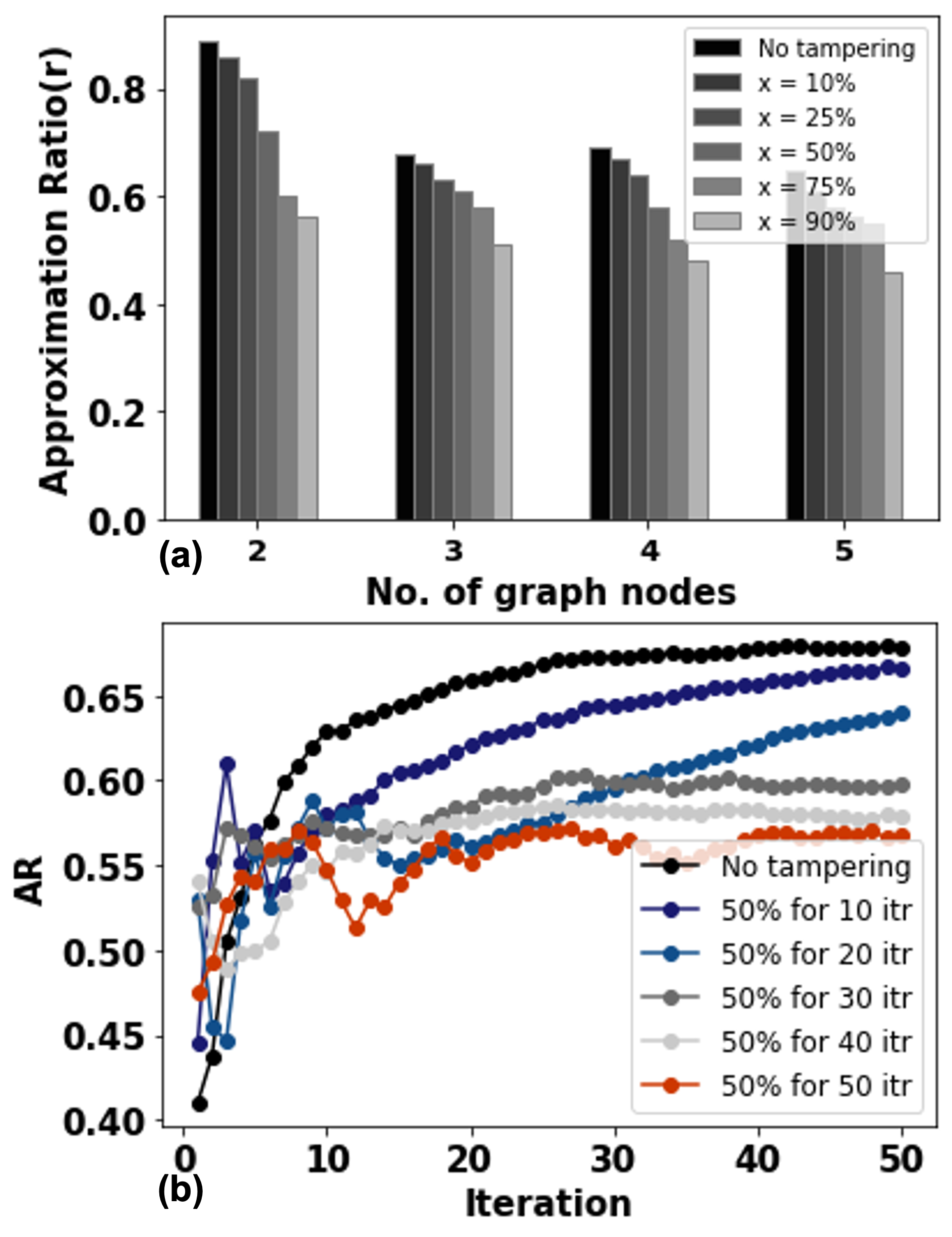}
    \caption{a) Approximation ratio (r) variation for different graph sizes when run on tampered fake$_-$montreal hardware with different parameter tampering error (x\%) for 50 iterations. b) AR variation of a 4-node graph with varying duration (in terms of iterations) of parametric tampering (x=50\%) when run on tampered fake$_-$montreal hardware. }
    \label{5}
\vspace{-4mm}
\end{figure}

\subsubsection{Tampering of parameters }

In this case, the attack will occur on the parameters of an iterative algorithm (such as, QAOA) before executing the interactive program on the quantum computer (Fig. \ref{3} attack model 2). The aim of adversary will be to prolong the convergence time and/or degrade the quality of solution. We investigate two distinct parameter attack situations: (a) The adversary adds some error to the parameter in all iterations. The following equation is used to model the tampered parameters for such scenario:

\begin{equation}\label{eq:er}
    (\gamma_-t, \beta_-t)   = (\gamma, \beta) \pm (x/100) * (\gamma,\beta)
\end{equation}
where $(\gamma, \beta)$ are parameters, ($\gamma$$_-$t, $\beta$$_-$t) are tampered parameters and $x$ is the tampering error.

(b) The adversary introduces significant error in the parameters during the first few iterations.

\section{Results and Analysis}

\subsection{Tampering Framework}

We model adversarial result tampering by introducing extra measurement error on the qubit lines i.e., while performing final measurement on a qubit, we flip the state of the qubit with probability $t$, which we refer to in the paper as the \emph{tampering coefficient}. In our work for simplicity  we introduce the bit flip error to a single bit line chosen randomly. We use IBM's fake backends to mimic real hardware and add this bit flip, tampering error.
For the adversarial parameter tampering we introduce \emph{tampering error} (x\%) in the parameters as per equation \ref{eq:er}.  

\subsection{Benchmark and Simulator}

For simulations, we use IBM's open-source quantum software development kit (Qiskit) \cite{b20}. A Python-based wrapper is built around Qiskit to accommodate the proposed attack model. Our benchmark suite include iterative (i.e., hybrid classical-quantum) algorithm QAOA. We use it to solve an exemplary combinatorial optimization problem namely, MaxCut \cite{b19} to investigate the effects of adversarial tampering on the hybrid quantum-classical algorithms. The MaxCut problem involves identification of a subset S$\in$V such that the number of edges between S and it's complementary subset is maximized for a given graph G = (V, E) with nodes V and edges E. Using a p-level QAOA, an N-qubit quantum system is evolved with $H_C$ and $H_B$ p-times to find a MaxCut solution of an N-node graph Fig. \ref{2}. QAOA-MaxCut iteratively increases the probabilities of basis state measurements that represent larger cut-size for the problem graph. We use the fake provider module in Qiskit as noisy simulators (Fake$_-$Montreal (27 qubit)) to run our benchmarks. The fake provider module contains providers and backend classes. The fake backends are created using system snapshots to mimic the IBM Quantum systems. Important details about the quantum system, including coupling map, basis gates, and qubit parameters (T1, T2, error rate, etc.), are contained in the system snapshots.

\begin{figure}
    \centering
    \includegraphics[width= 3.4in]{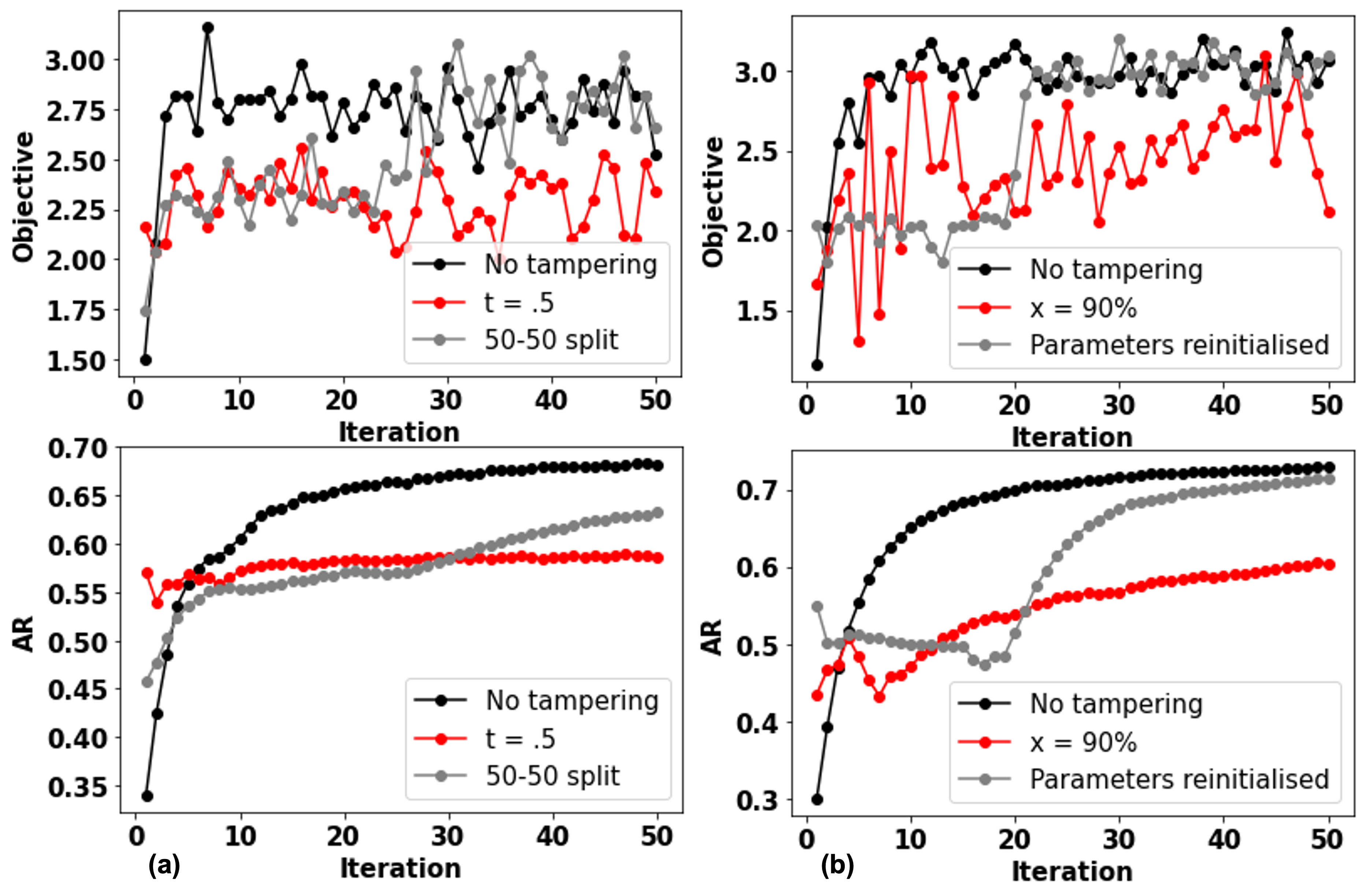}
    \caption{a) Effect of result tampering (t=.5) on the objective and AR for a 4-node graph over 50 iterations and proposed 50-50 iteration split defense. b) Performance of QAOA for 4-node graph where the adversary introduces significant error (x=90\%) during the first 15 iterations. As a defense against such an attack, we propose parameter re-initialization after initial few iterations. In this particular case we re-initialize parameters after 16$^{th}$ iteration. }
    \label{6}
\vspace{-4mm}
\end{figure}

\subsection{Simulation and Analysis}

For the sake of simplicity, we focus on MaxCut on unweighted d-regular node graphs (UdR), where each vertex is connected to only adjacent vertices. We use the approximation ratio defined in equation\ref{eq:AR} as the performance metric for QAOA. We run QAOA for each node graph ten different times and report the average values for the approximation ratio (r). The greater the r value, the better the performance. Ideally, the performance of QAOA can improve as p increases, with r $\rightarrow$ 1 when p $\rightarrow$ $\infty$. For our simulations, we run QAOA for maxcut on U2R, U3R, U4R, and U5R graphs (for p = 1) to investigate the effects of adversarial tampering on quantum-classical hybrid algorithms.
For the proposed result tampering model Fig. \ref{4}a shows the variation in AR for QAOA solving maxcut for various graph nodes. In each case, we run QAOA (p=1) for 50 iterations (with 50 shots/iteration). We note max $14\%$ (average $8\%$) and  max $30\%$(average $25\%$)  reduction in AR for t=0.1 and t=0.5, respectively. Fig. \ref{4}b depicts the variation in AR with the number of iterations for a 4-node graph run on tampered hardware with varying degrees of tampering. We see a decrease in approximation ratio for a fixed tampering constant when the number of iterations available to users for the tampered hardware is limited. When we run QAOA for 10 iterations rather than 50, AR degrades by 10$\%$ for $t=0.1$ and 25$\%$ for $t=0.5$, indicating that when run on tampered hardware, the performance of the hybrid-classical algorithm QAOA is sensitive to the number of available iterations. However, for tamper-free hardware, reducing the number of iterations from 50 to 10 results in a marginal ($2\%$) decrease in AR. As the probability of bit-flip error increases from $t=0.1$ to $t=0.5$, the adversarial tampering increases the overall qubit readout error, degrading the performance of the benchmark QAOA. The optimizer can no longer find the best parameters in the specified number of iterations. While $t=0.1$ incurs a cost of more iterations for the same performance, $t =0.5$ also results in a sub-optimal solution.

\begin{figure}
    \centering
    \includegraphics[width= 3.4in]{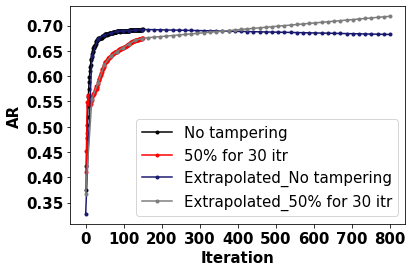}
    \caption{Approximation ratio (r) variation extrapolated to quantify iteration cost for a 4-node graph with parametric tampering (x=50\%) for the attack scenario in which the adversary targets parameters for the first 30 iterations.}
    \label{8}
\vspace{-4mm}
\end{figure}

We report decrease in AR of max $\approx 40\%$ (average $\approx 35\%$) across the various graph nodes between non-tampered and tampered hardware (for $x=90\%$) in the case of parameter tampering (Fig. \ref{5}a)) throughout the program execution time. The performance of QAOA for a 4-node graph with varying duration (in terms of iterations) of parametric tampering (x=50\%) is shown in Fig. \ref{5}b). We compare how the AR converges over 50 iterations if the adversary targets the parameters over a variable run duration. For the first few iterations of parametric tampering, the optimizer is able to recover the AR to approximately 97\% of its original value; however, for prolonged tampering (30-50 iterations), the AR does not recover. Fig. \ref{6}b) depicts the performance of a four-node graph in an attack scenario where the adversary introduces significant error in the parameters for the first 15 iterations ($x=90\%$). The optimizer returns the objective to the non-tampered value, but the AR does not recover over the 50 iterations. Because the optimizer receives the cost function generated by the adversary's tampered sub-optimal parameters, the number of iterations required to achieve the performance level of tamper-free hardware increases, resulting in lower performance for the same number of iterations when compared to tamper-free hardware. The figure \ref{8} shows how tampering affects the number of iterations required to achieve comparable QAOA performance. These figures are derived by extrapolating the 4-node QAOA performance for a 50\% parameter tampering error rate. In this scenario, the adversary tampers with the parameters for the first 30 iterations. To achieve comparable performance to a tamper-free hardware, the user incurs a minimum cost of 20X higher iteration.

\section{Proposed Defenses}

\subsection{Basic Idea and Assumptions} 

We assume that at least one of the $n$ hardware options available to the user is trustworthy, i.e. tamper-free. Users may also use the services of multiple untrustworthy cloud vendors, each with their own tampering model. The user, on the other hand, is unaware of the distinction between tampered and tamper-free hardware, as well as the adversarial tampering model. We propose splitting iterations either equally or intelligently on available hardware to mitigate the effects of adversarial tampering (from different trusted and untrusted vendors). For example, one might assume that hardware from well-known vendors AWS or IBM is trustworthy, whereas hardware from a less-known vendor X (which could be located in an untrustworthy country) is untrustworthy.

\begin{table}[]
    \centering
    \caption{AR vs Tampering (Iterations=50, Split=50:50); HW$_-$t(x/t) denotes tampered results for result tampering (modeled by varying t) and parametric tampering(modeled by varying x).}
    \begin{tabular}{cccccccc}
    \hline
      &  &  & Graph nodes &  &\\ 
     \hline
      &  & 2 & 3 & 4 & 5 \\ 
     \hline
    $t=0.1 /$ & HW & .84 & .67 & .68 & .68 \\
     $x=10\%$& HW$_-$t & .78/.81 & .6/.66 & .65/.67 & .65/.61 \\
     & Split & .81/.82 & .61/.66 & .67/.67 & .65/.62 \\
     \hline
    $t=0.2 / $ & HW & .84 & .67 & .68 & .68  \\
     $x=25\%$& HW$_-$t & .76/.79 & .56/.63 & .61/.64 & .58/.58 \\
     & Split & .78/.8 & .61/.65 & .65/.66 & .62/.61 \\
     \hline
    $t=0.3 / $ & HW & .84 & .68 & .68 & .68 \\
     $x=50\%$& HW$_-$t & .67/.72 & .54/.61 & .57/.58 & .56/.56 \\
     & Split & .7/.79 & .59/.63 & .61/.62 & .63/.6 \\
     \hline
    $t=0.4 /$ & HW & .84 & .66 & .68 & .68 \\
     $ x=75\%$& HW$_-$t & .57/.6 & .53/.58 & .54/.52 & .55/.55 \\
     & Split & .65/.75 & .6/.63 & .58/.61 & .63/.6 \\
     \hline
    $t=0.5 /$ & HW & .84 & .68 & .68 & .68 \\
    $ x=90\%$ & HW$_-$t & .5/.56 & .5/.51 & .5/.48 & .55/.46 \\
     & Split & .62/.64 & .59/.6 & .57/.58 & .6/.62 \\
     \hline
    $Avg AR\% \uparrow$ & & 10/11 & 10/12  & 9/12  & 11/15  \\
      \hline
     
     \end{tabular}
    \label{tab:1}
\end{table}

\subsection{Dividing Iterations Equally}

Without incurring any computing overhead, the user can distribute the iterations uniformly across the available hardware (assuming the hardware are homogeneous and queuing delays are identical). Assume the user has access to two different service providers' hardware, HW1 and HW2. HW2 is prone to tampering. The user must run a program P1 with a set number of iterations (an assumption, the program executes till convergence in reality). If he runs the program completely on  HW2, the results received will be tampered and unreliable. We propose three methods for the user to distribute computation among the various hardware: a) running half of the iterations on HW1, obtaining the parameters and using these parameters as the starting point for HW2 or vice versa, b) alternating iterations between HW1 and HW2, executing one iteration on HW1, followed by another on HW2, and so on. The proposed heuristics will make the results more resilient, reducing the likelihood of tampering by the adversary. These can be generalized to a scenario with n available hardware options.

\begin{table}[]
    \centering
    \caption{Intelligent iteration distribution : Identifying tampered/bad hardware (Iteration/run = 5) ; HW$_-$t(x/t) denotes tampered results for result tampering (modeled by varying t) and parametric tampering(modeled by varying x).}
    \begin{tabular}{ccccccc}

     \hline
      & & & Approximation ratio (r) \\ 
     \hline
    $t=0.1/$ &  $HW$ &  Run 1 & .61 \\
     $x=10\%$      &          & Run 2 & .59\\
           & $HW_-t$  & Run 1 & .56/.5 \\
           &          & Run 2 & .52/.48 \\
     \hline
     
    $t=0.2/$ &  $HW$ &  Run 1 & .62 \\
    $x=25\%$       &          & Run 2 & .63\\
           & $HW_-t$  & Run 1 & .53/.5 \\
           &          & Run 2 & .5/.54\\
     \hline
    $t=0.3/$ &  $HW$ &  Run 1 & .68 \\
    $x=50\%$       &          & Run 2 & .62 \\
           & $HW_-t$  & Run 1 &  .51/.47 \\
           &          & Run 2 & .50/.52 \\
     \hline
    $t=0.4/$ &  $HW$ &  Run 1 & .62 \\
     $x=75\%$      &          & Run 2 & .64 \\
           & $HW_-t$  & Run 1 & .49/.59 \\
           &          & Run 2 & .5/.47 \\   
     \hline
    
    $t=0.5/$ &  $HW$ & Run 1 & .62 \\
    $x=90\%$       &          & Run 2 & .66 \\
           & $HW_-t$  & Run 1 & .48/.48 \\
           &          & Run 2 & .49/.51 \\
     \hline
     \end{tabular}
    \label{tab:2}
\end{table}

\begin{figure}
    \centering
    \includegraphics[width= 3.4in]{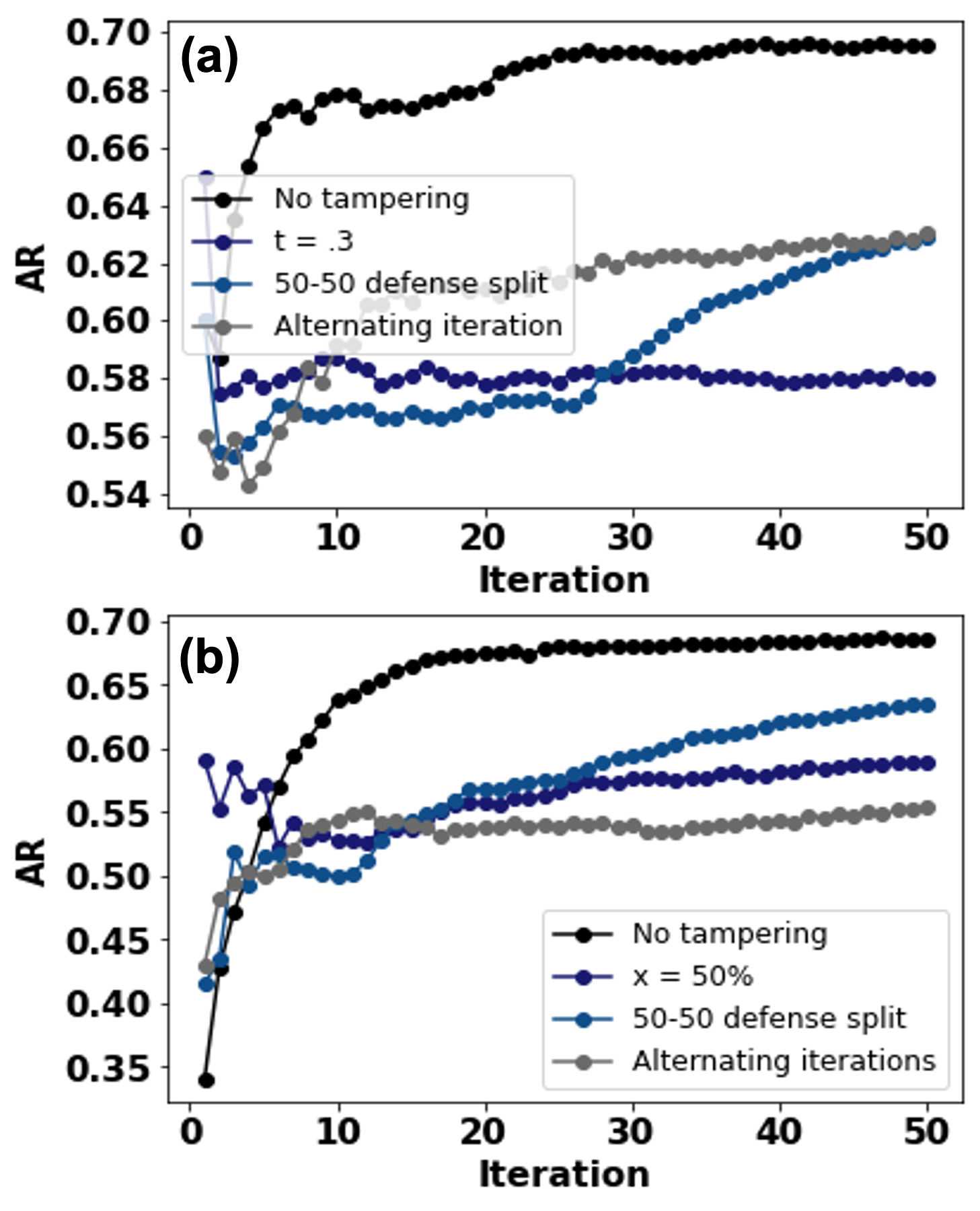}
    \caption{a) Defense against result tampering (t=.3) using proposed 50-50 iteration split and alternating iteration heuristic for a 4-node graph over 50 iterations. b) Defense against parameter tampering (x=50\%) using proposed 50-50 iteration split and alternating iteration heuristic for a 4-node graph over 50 iterations.}
    \label{7}
\vspace{-6mm}
\end{figure}

\subsection{Adaptive Iteration Allotment}

The iterations can also be distributed intelligently and adaptively by the user. This can be accomplished by running a few initial iterations on all available hardware, comparing the results, conducting majority voting, and then running the remaining iterations on the more reliable hardware.

\subsection{Re-initializing the parameters}

For the attack scenario in which the adversary introduces significant error in the parameters during the first few iterations prolonging the convergence time, we propose re-initialization of the parameters after a few initial iterations to recover the original program performance partially or fully. The user however, bears the cost of the repeated iterations.

\subsection{Simulation and Analysis}

Fig \ref{7} depicts the impact of the proposed 50-50 iteration split defense against the proposed attack models. Fig \ref{7}a We observe improvement in AR over iterations for result tampering model (t=.3)  by running half of the iterations on tampered hardware followed by other half on non-tampered hardware and also by alternating the iterations between the two hardware. However, the proposed alternating iteration heuristic does not work well for the parameter tampering model Fig \ref{7}b; we see comparable or worse performance when the program is run solely on tampered hardware. Hence, for our future simulations and experiments, we simply use the heuristic of running half of the iterations on tampered hardware, obtaining the parameters, and using these parameters as the starting point for tamper-free hardware, or vice versa for distributing the computation . Table \ref{tab:1} shows the improvement in AR with t, assuming one tampered (HW$_-$t) and one tamper-free of hardware (HW). We run 25 iterations (50 shots/iteration) on HW (fake$_-$ montreal) out of a total of 50 iterations, extract the parameters $(\gamma, \beta)$ after 25 iterations, and use them as a starting point for parameter optimization in HW$_-$t for another 25 iterations. We observe AR improvement for various levels of tampering. For $t=0.5$ and $x=90\%$, we report the maximum improvement in AR 33$\%$ (and 15$\%$ on average across various graph sizes.)

The performance of QAOA for a 4-node graph where the adversary introduces significant error (x=90\%) during the first 15 iterations is shown in Fig \ref{6}b. We propose parameter re-initialization after the first few iterations as a defense against such an attack. In this specific case, the user re-initialize parameters after 16$^{th}$ iteration. The user recovers the QAOA performance to nearly 99\% of its original value, which would have been limited to approximately 77\% if the countermeasure had not been used. In general the user can re-initialize the parameters at the cost of a few iterations to counteract such adversarial tampering.

A sample simulation of how a user can determine the tamper-free hardware and allocate the majority of iterations to that preferred hardware is shown in Table \ref{tab:2}. For a four-node graph, we run two 5-iteration (50 shot/iteration) runs on two different hardware HW (Fake$_-$montreal) and HW$_-$t (Fake$_-$montreal$_-$tampered). The simulations account for HW$_-$t's degree of tampering (by varying t from 0.1 to 0.5 for result tampering and x from 10\% to 90\% for parameter tampering). For each hardware, we compare the approximation ratio between the two runs. Higher AR hardware is better and more reliable. The user has the option of running the remaining iterations in HW only. As a result, the user can regain up to 95\% of the AR (when run solely on HW).
\vspace{-2mm}
\section{Discussion}

\textbf{Adversarial selection of parameters for tampering:}
To assess the efficacy of the attack model and defense, we only consider a single layer QAOA circuit with two parameters. For the sake of simplicity, we tampered with both parameters in our simulations. The adversary will have a large number of parameters available for tampering for other multi-layer circuits and iterative algorithms. The adversary can be selective and tamper with only one or a few parameters that are more important to the program's performance. Consider the following two parameters: alpha and gamma. Compared to gamma, alpha requires more iterations to stabilize around the optimum value. The adversary can be more selective and less invasive by only targeting alpha, imposing higher convergence time to the users.

\textbf{Adversarial selection of iterations for tampering:}
We considered either a continuous parameter tampering or tampering during the first few iterations. However, the adversary can tamper with the parameters at any time during the program. The adversary may also choose to tamper with the program's results near the end. However, for such a tampering model, the drop in performance will expose the adversary's tampering efforts. In such cases, the optimizer can be programmed from the user end to stop the program execution once the dip in parameters quantifying program performance is detected.
\vspace{-2mm}
\section{Conclusion}

We propose an adversarial attack by a less trustworthy third-party provider  on a hybrid quantum-classical benchmark (QAOA). Across various node graph sizes, we report an average reduction in AR of 2\%, 3\% (t=0.1, x=10\%) and 25\%, 28\% (t=0.5, x=90\%). We report a significant increase in the number of iterations required to achieve comparable performance to a tamper-free hardware ($\approx$20X more). To ensure trustworthy computing using a mix of trusted and untrusted hardware, we propose distributing the total number of iterations available to the user among various hardware options. We see a max AR improvements of up to 45\% and average improvement of 20$\%$ across the various defense heuristics proposed. Our proposed heuristics (50-50 iteration distribution, run-adaptive iteration distribution, and parameter re-initialization) reduce the adversary's tampering (result/parameter) efforts, improving the quantum program's resilience.

\section{Acknowledgments}
We thank Dr Rasit Onur Topaloglu (IBM Corporation) for useful discussions. The work is supported in parts by NSF (CNS-1722557, CNS-2129675, CCF-1718474, OIA-2040667, DGE1723687, DGE-1821766, and DGE-2113839) and seed grants from Penn State ICDS and Huck Institute of the Life Sciences.
\bibliographystyle{ACM-Reference-Format}
\bibliography{ref}

\end{document}